\documentclass[journal]{IEEEtran}

\usepackage{array}
\usepackage{graphicx}
\usepackage{amsmath,amsfonts,amssymb,bm}
\usepackage{xr}
\usepackage{threeparttable}

\usepackage[acronym,nomain]{glossaries}
\setacronymstyle{long-short}
\newacronym{ac:dcdrs}{DCDRS}{Drop-Coating Deposition Raman Spectroscopy}
\newacronym{ac:diw}{DIW}{Deionized Water}
\newacronym{ac:hsv}{HSV}{Herpes Simplex Virus}
\newacronym{ac:hcpcf}{HC-PCF}{Hollow-Core Photonic Crystal Fiber}
\newacronym{ac:ipa}{IPA}{Isopropyl Alcohol}
\newacronym{ac:lcw}{LCW}{Liquid Core Waveguide}
\newacronym{ac:na}{NA}{Numerical Aperture}
\newacronym{ac:patti}{PATTi}{Photonic Adhesive Tip Tamping}
\newacronym{ac:pdms}{PDMS}{Polydimethylsiloxane}
\newacronym{ac:pc}{PC}{Principal Component}
\newacronym{ac:pcr}{PCR}{Principal Component Regression}
\newacronym{ac:press}{PRESS}{Predicted Residual Sum of Squares}
\newacronym{ac:rms}{RMS}{Root Mean Square}
\newacronym{ac:rss}{RSS}{Residual Sum of Squares}
\newacronym{ac:sers}{SERS}{Surface-Enhanced Raman Spectroscopy}
\newacronym{ac:tct}{TCT}{Teflon Capillary Tube}
\newacronym{ac:tir}{TIR}{Total Internal Reflection}

\begin{document}

\title{Enhanced Sensitivity for Quantifying Disease Markers via Raman and Machine-Learning of Circulating Biofluids in Optofluidic Chips}


\author{Emily~E.~Storey,
	    Duxuan~Wu,
	    Amr~S.~Helmy,~\IEEEmembership{Senior Member,~IEEE,}

\thanks{E.E. Storey and A.S. Helmy are with the Department of Electrical and Computer Engineering, University of Toronto, Toronto, Ontario, Canada.}
\thanks{Corresponding Author E-mail a.helmy@utoronto.ca}%
\thanks{D. Wu was with the Department of Electrical \& Computer Engineering, University of Toronto, Toronto, Ontario, Canada.}}

\IEEEoverridecommandlockouts
\IEEEpubid{\makebox[\columnwidth]{\begin{minipage}{\columnwidth}\copyright2021 IEEE. Personal use of this material is permitted. Permission from IEEE must be obtained for all other uses, in any current or future media, including reprinting/republishing this material for advertising or promotional purposes, creating new collective works, for resale or redistribution to servers or lists, or reuse of any copyrighted component of this work in other works.\hfill\end{minipage}} \hspace{\columnsep}\makebox[\columnwidth]{ }}
\maketitle
\IEEEpubidadjcol
\IEEEpubidadjcol

\begin{abstract}
	We demonstrate novel instrumentation for spontaneous Raman spectroscopy in biofluids, enabling development of a portable, automated, reliable diagnostics technique requiring minimal operator expertise to quantify disease markers. Label-free Raman analysis of biofluids at physiologically-relevant sensitivities is achieved using a microfluidic-embedded liquid-core-waveguide augmented with a unique circulation approach: thermal damage and spectrum variance is minimized, eliminating conventional limits on integration time for excellent signal-to-noise ratio and temporal stability. Machine-learning then optimizes spectrum processing, yielding quantitative results independent of end-user proficiency. Sub-mM accuracy is achieved in solutions of both high and low turbidity, surpassing the sensitivity of previous techniques for analytes with a small scattering cross-section, such as glucose. We attain a new record for label-free glucose measurements in an artificial whole-blood, achieving an accuracy up to 0.14 mM, well-exceeding the 0.78 mM accuracy required for diabetic monitoring, establishing our technique's potential to significantly facilitate portable Raman for complex biofluid analysis.
\end{abstract}


\IEEEpeerreviewmaketitle

\section{Introduction}

	\IEEEPARstart{O}{n-site} rapid-turnaround health monitoring is increasingly in demand to maximize efficiency and minimize patient stress. Biofluids are a promising means for non-invasive diagnostics, faciliting routine monitoring and replacing invasive investigative procedures which may entail complications, thus enhancing patient quality of life. To this end we seek an analytic system which offers enhancement in sensitivities reaching physiologically relevant values; when developed, this system has diagnostic potential in a range of biofluids which each offer unique insight into a patient's state of health. 
	
	Four factors will play a key role in the success of a rapid-turnaround health monitoring system: portability, non-invasive sample collection, no specimen pre-treatment, and reliable automated diagnostic interpretation of the results. Diabetes management excellently illustrates the need for a system with these four features. Regular and accurate monitoring of blood glucose concentration is key to successful management, yet many find finger pricking, the traditional standard, to be painful. Non-invasive biofluids, such as tears, can serve as an alternate monitoring fluid, increasing compliance to decrease mortality \cite{Lane2006}. These four factors are equally essential for patients dealing with a sudden illness onset: rapid diagnosis can have a drastic influence on  survivability, both to the individual and to the community at large depending on pathogenicity. This situation is demonstrated during the Covid-19 pandemic: as under-funded health systems struggle to maintain high testing rates, the virus surges where there is significant delay between disease contraction and diagnosis \cite{Bennett2021}.
	
	Several techniques have been proposed in recent years as alternate biofluid sensing platforms \cite{Li2021}. Paper-based devices, such as lateral-flow assays, have become widespread in part due to their ease of use and cost-effectiveness. These devices are highly specific, requiring multiple tests to detect multiple analytes - an option that can be complicated if further sample is unavailable \cite{Wang2016}. A second alternate technique is a flexible biosensor for tear glucose measurement, which shows promising correlation between output current and glucose levels \cite{Iguchi2007}. Devices of this type can arguably satisfy the four factors we desire, but there is high potential for interference by other electroactive species. The aforementioned devices must be tailored to each particular analyte. While the sample itself does not require pre-treatment, the measurement system must be reconfigured to detect different targets, reducing applicability towards a range of diseases. Optical methods do not suffer this drawback.
	
	Non-invasive biofluid analysis is readily accomplished by Raman spectroscopy due to its excellent chemical specificity. Spontaneous Raman is notorious for a faint scattering signal, however, thus not serving our desire for a pre-treatment-free system. Amplification solutions such as high-powered laser sources are damaging to biological specimens, while pre-treatments such as \gls{ac:sers}, or \gls{ac:dcdrs} require specially-prepared substrates or functionalization which is dependent on the reagents involved. It is prohibitive to yield consistent spectra without highly-trained personnel, making these methods unsuitable for straightforward diagnostic interpretation. Spontaneous Raman with \gls{ac:lcw} optofluidics enhances the collection efficiency without these specimen modifications. This technique has shown that it provides superior sensitivity, enabling native-state analysis of compounds which would not otherwise be possible using Raman. We have previously demonstrated the use of \gls{ac:lcw} Raman as a probe to monitor growth of ZnO nanoparticles, and in the world-first detection and characterization of thiol-capped CdTe Quantum Dots, with zero impact on their molecular properties \cite{Irizar2008, Mak2011}. However these prior optofluidic Raman demonstrations, alongside more recent related literature, are not practical for implementation outside of a laboratory environment: they require that the analyte is contained in a bulky apparatus, or is exposed to the environment, posing a risk of contamination and exposure to pathogens \cite{Persichetti2015,Yan2017}. 
	
	We demonstrate here a system which eliminates these hurdles, using optofluidic Raman to achieve comparable signal enhancement to that obtained by \gls{ac:sers} in a manner which makes portable implementation and straightforward pre-treatment-free diagnostics a realizable goal. The complete fluid-based monitoring system which we demonstrate is ready to integrate with portable Raman spectroscopy for analysis of non-invasively-collected biofluids at physiologically-relevant sensitivity of detection. We achieve such sensitivities using continuous-flow optofluidics, minimizing thermal damage to the sample and thereby eliminating limits on integration time. The hardware which we demonstrate serves three of the key factors we desire: portability, non-invasive sample collection, and no specimen pre-treatment. The final factor, reliable automated diagnostic interpretation, is addressed through machine-learning. Variation in biological samples is inherent even from one individual: these variations are reflected in Raman spectra and it is crucial to compensate for with a high degree of reproducibility in order to accurately assess one patient relative to a normal standard \cite{Sen1980,Chiappin2007}. Spectrum preprocessing techniques are common to mitigate these variations, but the decision of which specific technique is most suitable requires manual intervention, and thus a predictive model's robustness will vary with the operator \cite{Zeaiter2005}. Thus we introduce a machine-learning algorithm for preprocessing technique selection, achieving user-independent spectrum optimization.
	
	The different components of this system, once integrated, provide previously untenable performance from a single device, for robust automated diagnostics capability. We demonstrate excellent chemical specificity, readily adaptable to quantitatively analyze a range of fluids with accuracies at or exceeding the $\mu$M level.

\section{Methods}

	\subsection{Experimental Design}
		All spectra were collected on a Horiba HR800 spectrometer equipped with three excitation laser wavelengths: 488.2 nm (3.5 mW at the sample), 632.8 nm (10 mW), and 785.0 nm (30 mW).
		
	\subsubsection{Microfluidics}
\begin{figure}[!t]
			\centering\includegraphics[width=\linewidth]{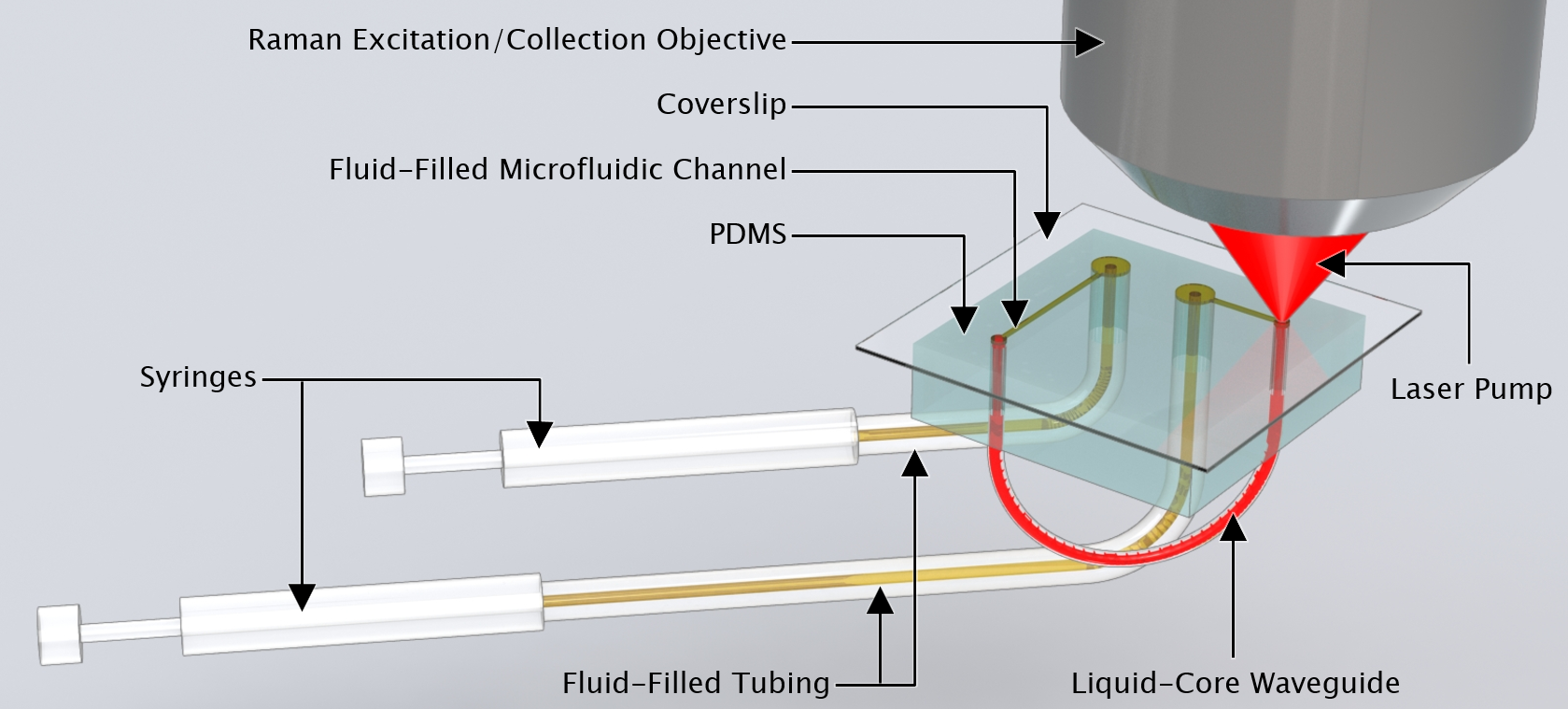}
			\caption{Integrated-waveguide microfluidic design. Microfluidic chip consists of a layer of \acrfull{ac:pdms} patterned with fluidic channels and irreversibly bonded to a glass coverslip. Fluidic sample is depicted as dark yellow in tubing and microfluidic channel; inside the waveguide it is depicted in red, illustrating the co-location of sample and exciting laser source. Direct tubing connections allow a syringe pump to supply differential pressure between two syringes, such that the biofluid never leaves the closed-loop system and can be collected after measurement, eliminating environmental exposure and sample wastage.}
			\label{fig:MFSchematic}
		\end{figure}

		Fluidic sample circulation to facilitate portable and pre-treatment-free sample analysis is accomplished via an integrated microfluidic waveguide-on-chip. This design builds upon previous work in which the waveguide facet is exposed to air and fluid is introduced to the waveguide core by capillary action. Containing the sample and \gls{ac:lcw} in a microfluidic device eliminates sample evaporation at the facet, maintaining optimal optical coupling. A dual-syringe system, one syringe to dispense and one to collect via differential pressure, interfaces with the microfluidic device to circulate sample through the waveguide during a measurement. This continuous-flow eliminates previous limits on integration time for delicate biological samples which may crystallize or denature in the presence of high-intensity laser sources, as the volume of sample which resides in the focal volume is continually refreshed, thus dispersing the thermal load. This microfluidic and syringe system additionally eliminates waveguide length limitations. Within the waveguide, laser and sample are co-located throughout the full length: increasing the total volume of sample which interacts with the laser increases the number of scattering events and thereby the quality of spectrum \cite{Eftekhari2011}. Figure \ref{fig:MFSchematic} details this configuration. 
		
		Microfluidic devices consist of \gls{ac:pdms} (Sylgard-184) adhered to a glass coverslip. Biopsy punches are used to bore cylinders in cured \gls{ac:pdms} for press-fit \gls{ac:lcw} and tubing connections. 
		
		Two different \glspl{ac:lcw} have been tested in our device: \glspl{ac:tct} (AF 2400, Biogeneral, Inc.) and \glspl{ac:hcpcf} (HC-800 and HC-1060, NKT Photonics). \Glspl{ac:tct} provide Raman enhancement solely through \gls{ac:tir}. \Glspl{ac:hcpcf} additionally enhance the Raman collection efficiency via the photonic bandgap effect \cite{Mak2013}, resulting in an enhancement upwards of two orders of magnitude relative to the signal obtained from a \gls{ac:tct}, and three orders of magnitude relative to the Raman spectrum of a droplet of fluid exposed to air, as shown in Supporting Information, figure \ref{fig:TCTvsDrop}. Panel B of the same figure contrasts the enhancement provided by each waveguide for a single glucose Raman mode, located at 1127 cm$^{-1}$, as a function of concentration, normalized to exposure time. It is not until optofluidics is merged with continuous circulation that we are able to integrate indefinitely and obtain spectra of excellent signal to noise ratio for optimized concentration determination.
		
		Optofluidic integration of \glspl{ac:hcpcf} requires that the fluid is restricted from entering the microstructured cladding of both fiber facets in a manner which does not compromise coupling to the fiber core. This is accomplished in a two-step process, dubbed \gls{ac:patti}. A thin layer of ultraviolet-curable adhesive is applied across the tip of the fiber, while a burst of air clears adhesive from the central core. Specialized equipment is minimal: a micromanipulator, microscope, syringe pump, and ultraviolet source. Full procedure is included in Supporting Information, figure \ref{fig:supp-patti}. \Gls{ac:patti} has a yield over 80\% and allows us to routinely achieve adhesive-sealed lengths below 20 $\mu$m at both ends of the fiber. 
		
		Length of \glspl{ac:lcw} in all experiments is 25 mm. Exposure times ranged from 3 to 480 seconds per detector window, for a total spectrum collection time up to 2 hours. Species which are highly Raman active, suspended in a low-scattering medium and excited with a high-energy laser required very short exposure times. When the opposite situation occurs, Raman spectra of similarly excellent quality can likewise be collected by extending exposure times to compensate. 
		
		The samples described in this experiment were circulated at rates between 0.05 and 0.33 $\mu$L/min. For \glspl{ac:tct}, this means the fluid within the waveguide core is refreshed every 2 to 15 minutes, and for \glspl{ac:hcpcf}, every 1.5 to 10 minutes. The lower limit for flow rate was selected such that the fluid within the waveguide core is being refreshed on a reasonable time scale within which thermal damage is unlikely to occur. The upper limit was selected to mitigate pressure drop within the device which may cause press-fit tubing connections to disconnect. Within the quoted range of flow rates, the fluid velocity within the waveguide core did not present a noticeable influence on the quality or intensity of a Raman spectrum.
		
	\subsubsection{Artificial Biofluids}
		Two artificial biofluids are prepared for this study: one to mimic human tear fluid and one which emulates the high-scattering whole human blood. 
		
		Samples to mimic human tears were comprised of lysozyme and glucose dissolved in \gls{ac:diw}, each at physiologically relevant levels. D-(+)-Glucose (G8270) and Lysozyme from chicken egg white (62970) were purchased from Sigma-Aldrich. 40 unique samples were prepared with analyte distribution primarily concentrated within the range of diagnostic relevance, down to a minimum sensitivity of 0.17 mM lysozyme and 1.4 mM glucose \cite{Sen1980,VanHaeringen1977,Eylan1977}. Maximum analyte concentrations in our samples exceeded the upper bounds for diagnostic relevance. These elevated concentrations allow us to confirm the presence of known Raman modes for each analyte, particularly glucose due to its small scattering cross-section. This combination of components was selected to provide a spectrum which allows us to discern the concentration of the desired analyte under varying background conditions. This is a necessary first step upon which future work will build, as the Raman spectra of human biofluids will contain other components which complicate the baseline spectrum against which we seek to measure the analyte of interest.
		
		A 20\% Intralipid fat emulsion (Sigma-Aldrich I141) acts in stead of whole human blood; its scattering coefficient is a good match to that of whole blood and it requires no special handling or disposal \cite{Bosschaart2014,Michels2008}. These samples were prepared to test the limitations of detecting a species with weak Raman signature in a highly scattering medium. 15 binary solutions were prepared between 0 mM and 138.8 mM glucose, primarily concentrated in the range 278 $\mu$M to 2.78 mM. 

	\subsection{Machine-Learning Model Optimization}\label{sec:ml}
	
		To achieve reliable automated diagnostics from Raman spectra, we have previously introduced a machine-learning algorithm to optimize spectrum processing for analysis \cite{Storey2019}. Here we build upon those methods: the parameter which selects a solution for constituent concentration determination is tested against alternates, to determine which parameter optimizes the solution and minimizes machine learning error. 
		
\begin{figure}[!t]
			\includegraphics[width=\linewidth]{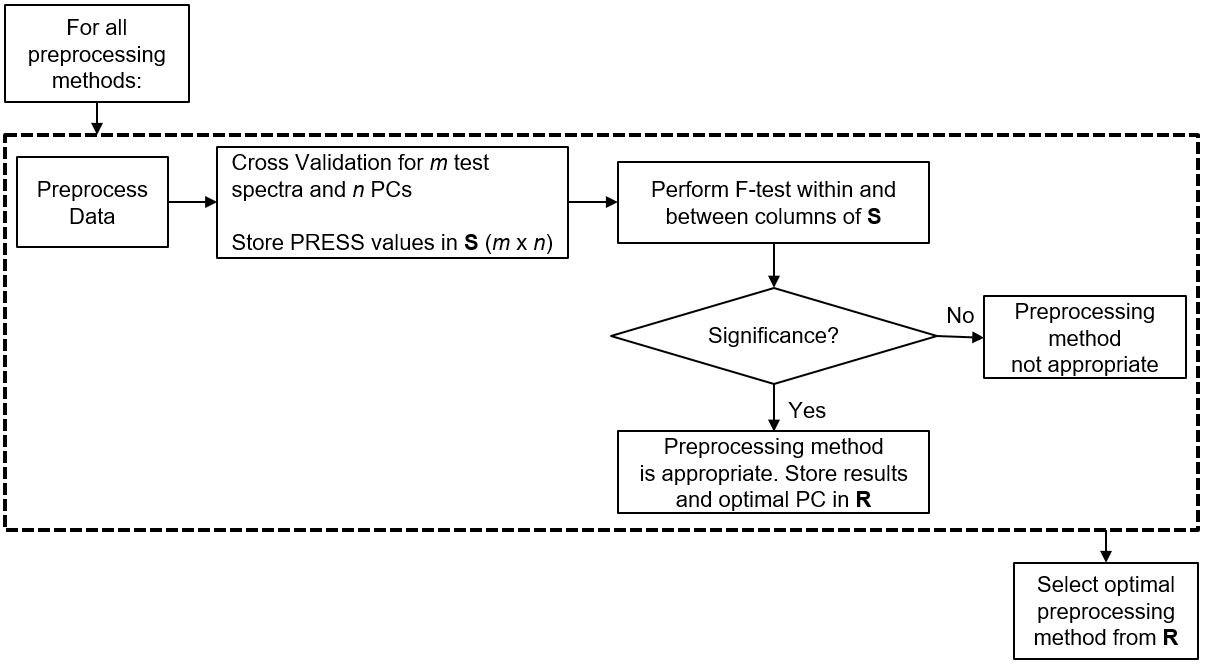}
			\centering
			\caption{Machine-learning algorithm. All spectra in a set are preprocessed and cross-validated to determine statistical significance of cross-validation. Only those methods which demonstrate a significant variation in Predicted Residual Sum of Squares (PRESS) as the number of Principal Components (PCs) is altered are passed on to be considered for composition prediction via Principal Component Regression (PCR)}
			\label{fig:ML}
		\end{figure}

		\Gls{ac:pcr} is used for predictive analysis. As the highest-order \glspl{ac:pc} inevitably represent noise and are of no value to a predictive model, the number of \glspl{ac:pc} in the model must be carefully chosen not to exclude subtle spectrum details in relatively high-order \glspl{ac:pc} which positively contribute to prediction accuracy. We accomplish this by performing cross-validation on a training data set for preprocessing methods under consideration. Results are then stored in a \gls{ac:press} matrix for variance analysis. An F-test (significance level $\alpha$ = 0.05) assesses the variability in this matrix to determine if improvement in prediction is statistically significant or due to sampling. This process rejects preprocessing methods which demonstrate chance correlation and is presented in figure \ref{fig:ML} \cite{Storey2019}.
		
		Utilizing only those preprocessing methods which display statistical significance provides a numerical measure of assurance that the model is robust and will perform similarly well when applied to new spectra. Throughout this discussion we shall refer to this as the current method, as it has been successfully utilized in previous work \cite{Storey2019}. However, a model constructed in complete absence of user input will only be as robust as the selection of optimal \gls{ac:pc} allows. To build upon the work in \cite{Storey2019} we aim to statistically optimize the indicating factor which identifies optimal \gls{ac:pc}, comparing the current method to 10 other indicators. To do this we form a linear regression model between the indicating variables and \gls{ac:rss}. A linear fit is performed and R$^2$ and median error values from this fit are used to guide our assessment of the indicator(s) which optimize \gls{ac:pc} selection. Full description of indicator variables is included in Supporting Information table \ref{tab:indicatingVariables}.

\section{Results}

	We demonstrate in this section the design of a label-free user-independent quantitative analysis system for biofluids, providing proof of concept for reliable detection and quantification of disease markers in a portable automated setting. This is accomplished with a microfluidic device with embedded waveguide for closed-system continuous-flow and circulation of the solute under test. Microfluidics lends itself readily to portability, easily realizing a small-scale device with flexible configuration in which the sample is circulated for measurement. 
	
\begin{figure}[!t]
		\centering\includegraphics[width=\linewidth]{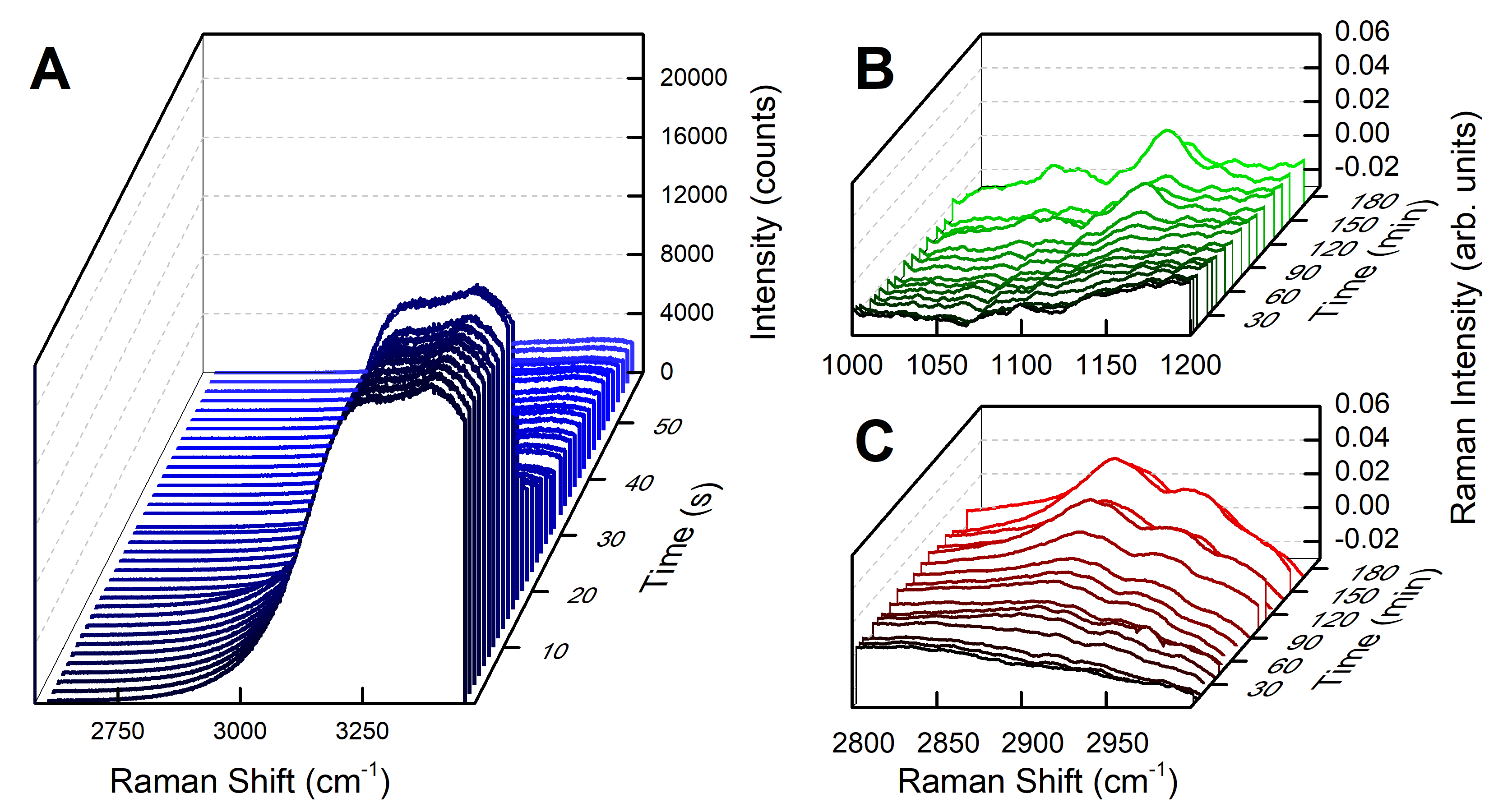}
		\caption{Temporal changes in Raman spectra of a static fluid measured using \acrfull{ac:hcpcf}. \textbf{A}: Fluid reservoir is removed, causing the signal intensity to degrade within seconds. \textbf{B} \& \textbf{C}: A solution of 0.5\% (m/v) glucose in deionized water is measured over several hours; glucose modes become increasingly prevalent relative to the spectrum at time 0 due to analyte deposition and crystallization, making the solution appear more saturated than its true concentration.}
		\label{fig:tempGlucose}
	\end{figure}

	This closed-system continuous-flow system, illustrated in figure \ref{fig:MFSchematic}, provides temporal stability to the Raman spectrum on two critical aspects. First, if the fluid reservoir is not sufficiently maintained, the sample disappears from the \gls{ac:hcpcf} facet in a matter of seconds. This case is illustrated in figure \ref{fig:tempGlucose} panel A in which a vial of sample (\gls{ac:diw}) is removed, causing signal intensity to drop by more than 80\%. Second, we directly inhibit structural or chemical changes to the biofluid over extended measurement periods by dispersing the thermal load. These modifications alter the resultant spectrum and thus our ability to reliably quantify. A sample case is illustrated in figure \ref{fig:tempGlucose}, panels B and C, in which the Raman spectrum of a static solution of glucose dissolved in \gls{ac:diw} (2.8 mM) is measured using an \gls{ac:hcpcf} as described in \cite{Mak2013}. Spectra presented are relative to the spectrum at time 0. Extended integration times are necessary to sufficiently resolve Raman modes of trace analytes, but spectra are not temporally stable in this configuration: glucose modes in the vicinity of 1127 cm$^{-1}$ and 2900 cm$^{-1}$ become increasingly prevalent. The result is an unacceptable ambiguity in interpretation for diagnostics.
	
	To demonstrate the claims of reliable automated diagnostics with no sample pre-treatment, two artificial biofluids are constructed to emulate the optical properties of tears and whole blood, representing examples of non-invasive and invasive bio-analysis, respectively. The accuracy and performance we achieve lays the groundwork to apply similar methods to a much wider array of fluids and diagnostically-relevant analytes beyond what is presented. These alternate biofluids include urine and saliva, where their potential will be explored in the Discussion. An essential component to accurate composition prediction for diagnostics is the selection of appropriate preprocessing treatment for raw data. We automate this selection, marrying \gls{ac:pcr} with a measure of statistical significance to indicate that the model's predictive capabilities are not due to chance correlation with training data. 
	
\begin{figure}[!t]
		\includegraphics[width=\linewidth]{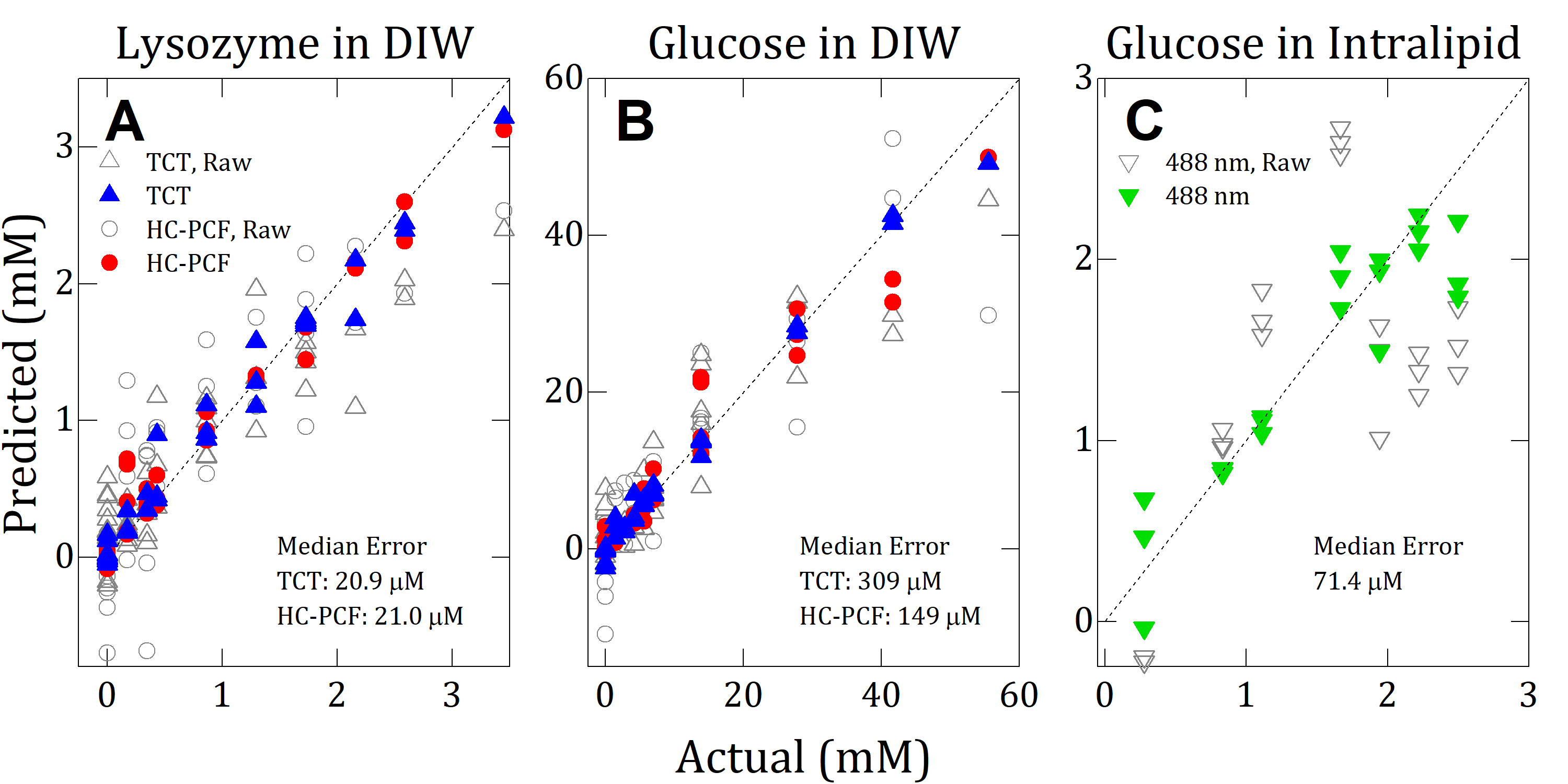}
		\centering
		\caption{Actual versus inferred concentrations, determined by our algorithm. Dashed diagonal line indicates perfect correlation in concentration; an open symbol indicates no preprocessing. \textbf{A} \& \textbf{B}: Glucose and lysozyme in a \acrfull{ac:diw} artificial tear solution, measured using either \acrfull{ac:tct} or \acrfull{ac:hcpcf}. \textbf{C}: Glucose predictions in an intralipid artificial whole-blood solution, using a 488 nm excitation laser and a HC-PCF.}
		\label{fig:All_1to1}
	\end{figure}

	Accurate measures of analyte concentration with respect to reference value are paramount, in particular for chemometrics relating to physiological metrics, to ensure that predictions will lead to appropriate treatment methods. At best, failure to accurately quantify may result in prolonged discomfort for a patient seeking to rectify an ailment. At worst, fatal failure to detect and enact treatment may occur. We present in figure \ref{fig:All_1to1} the estimated versus actual concentration of each analyte under test using our methods. Cases in which we have used machine-learning to select the optimal \gls{ac:pc} but applied no preprocessing are shown with open symbols. A dashed line indicates perfect correlation. 

	Significance of these results are presented with four metrics of accuracy in table \ref{tab:resultsAll}: 1) error in prediction with no preprocessing or machine-learning, 2) when a preprocessing method is not found to have statistical significance, 3) when a preprocessing method is statistically significant, and 4) optimal error in prediction amongst all statistically significant preprocessing methods. Optimal predictions automate the selection of number of \glspl{ac:pc} to form a predictive model, and all preprocessing techniques which display significance are manually selected amongst. No user intervention is required to determine statistical significance, and these methods repeatedly yield improved accuracy over those that do not display significance. 
	
	\begin{table*}[!t]
		\centering
		\caption{Error in prediction for analytes in two different simulated biofluids: artificial tears (\acrfull{ac:diw} containing glucose and lysozyme), and an artificial whole blood solution (intralipid containing glucose). Two liquid-core waveguides (\acrfull{ac:tct} and \acrfull{ac:hcpcf}) enhance the Raman collection efficiency. Excitation wavelength is 633 nm unless otherwise noted.}
		\label{tab:resultsAll}
		\begin{tabular}{>{\centering\arraybackslash}m{5cm}  >{\centering\arraybackslash}m{2.5cm} >{\centering\arraybackslash}m{2.5cm} >{\centering\arraybackslash}m{2.5cm} >{\centering\arraybackslash}m{2.5cm}}
			\hline 
			\textbf{Sample Description} & \textbf{Insignificant (mM)} & \textbf{Significant (mM)} & \textbf{Optimal (mM)} & \textbf{Raw Spectra (mM)}\\ 
			\hline
			Glucose in DIW, TCT 					& 4.96 $\pm$ 10.5 		& 2.64 $\pm$ 3.62 		& 0.309 $\pm$ 2.57		& 2.37 $\pm$ 3.63\\ 
			Glucose in DIW, HC-PCF 					& 5.07 $\pm$ 10.0 		& 1.93 $\pm$ 3.69 		& 0.149 $\pm$ 1.20		& 1.57 $\pm$ 4.84\\
			Lysozyme in DIW, TCT 					& 0.404 $\pm$ 0.646 	& 0.230 $\pm$ 0.305 	& 0.0209 $\pm$ 0.116 	& 0.234 $\pm$ 0.260\\
			Lysozyme in DIW, HC-PCF 				& 0.475 $\pm$ 0.583 	& 0.246 $\pm$ 0.255 	& 0.0210 $\pm$ 0.135 	& 0.312 $\pm$ 0.285\\
			Glucose in Intralipid, HC-PCF, 488 nm 	& 0.522 $\pm$ 0.541 	& 0.744 $\pm$ 0.466 	& 0.0714 $\pm$ 0.214 	& 0.719 $\pm$ 0.316\\
			Glucose in Intralipid, HC-PCF 			& 18.6 $\pm$ 24.6 		& 56.4 $\pm$ 43.9 		& 28.4 $\pm$ 27.5 		& 27.2 $\pm$ 30.7\\
			Glucose in Intralipid, HC-PCF, 785 nm 	& 15.5 $\pm$ 27.5 		& 21.6 $\pm$ 26.4 		& 4.54 $\pm$ 23.3 		& 23.7 $\pm$ 25.0\\
			\hline 
		\end{tabular} 
	\end{table*}

	\subsection{Biofluid Results}
	
	\subsubsection{Glucose and Lysozyme in Simulated Human Tear Fluid}
	
	Artificial tear solutions designed for this study are composed of \gls{ac:diw} with solutes glucose and lysozyme; here we shall discuss their respective potential means for non-invasive diabetes monitoring, and as a tear proteins biomarker for \gls{ac:hsv}. While the bulk of this paper is focused on applicability to glucose quantification for diabetes management, we include lysozyme and \gls{ac:hsv} determination for context of applicability towards other ailments. 
	
	Optimal \gls{ac:pcr} predictions for each analyte and sample, measured by both \gls{ac:tct} and \gls{ac:hcpcf}, are shown in figure \ref{fig:All_1to1} panels A-B; median errors of prediction are documented in table \ref{tab:resultsAll}.
	
	A glucose measurement device must predict concentrations within 20\% of reference values to be considered clinically accurate and free from potentially-fatal disease mistreatment \cite{Clarke1987}. Here we have achieved a median accuracy up to 298 $\mu$M, a 21\% error on the minimum sensitivity of our experiment (1.4 mM). While this value does not meet the threshold for clinical accuracy, we shall later demonstrate that our selection algorithm can be optimized to meet and exceed 20\% error. 
	
	\Gls{ac:hsv} is highly contagious and cannot be cured; as such, effective detection is paramount to preventing its spread, ensuring appropriate preventative measures are taken. Mean lysozyme level in tears of patients with \gls{ac:hsv} is 1.0 mM and without is 2.1 mM \cite{Eylan1977}. When measured using our devices, artificial tear solutions were detected with a minimum lysozyme sensitivity of 0.17 mM. This sensitivity of detection exceeds the minimum level at which lysozyme is expressed in human tears as a consequence of \gls{ac:hsv}. We observe a median accuracy as low as 42 $\mu$M amongst optimal methods processed using our statistical algorithm, well in excess of the accuracy required for \gls{ac:hsv} determination.
	
	\subsubsection{Glucose Predictions in Simulated Whole Blood}

	To establish proof of concept and test accuracy of detection for an analyte with a small Raman scattering cross-section within a high-scattering opaque artificial biofluid, we present results from samples composed of an intralipid fat emulsion and powdered glucose. Haemoglobin in whole blood is an excellent light scatterer and samples are typically centrifuged prior to analysis; these results demonstrate a facile method to compensate for traditional measurement difficulties, an essential precursor to pre-treatment-free blood diagnostics using Raman. 
	
	Minimum glucose sensitivity in these artificial whole-blood samples is 280 $\mu$M glucose. All spectra were collected using a \gls{ac:hcpcf} waveguide and excitation wavelengths of 633 nm and 785 nm. These wavelengths were chosen strategically such that a \gls{ac:hcpcf} (HC-800, NKT Photonics) guides their respective Raman shifts via photonic bandgap effect (50-1500 cm$^{-1}$ for 785 nm, and 2500-3750 cm$^{-1}$ for 633 nm). A third excitation wavelength, 488 nm, was used for a selection of low-concentration samples, providing Raman collection enhancement by \gls{ac:tir}. The combination of these three wavelengths is appropriate for the sample at hand (intralipid), but the same combination does not extend to a human whole-blood sample due to strong fluorescence in that region. Appropriate pairing of sample and Raman wavelength, as we have done here for intralipid, is a necessary precursor to characterization of biological samples in order to mitigate fluorescence. \Glspl{ac:hcpcf} assist to this end, as their narrow photonic bandgap restricts the spectrum propagating through the sample to those wavelengths within the bandgap-guided range, thus suppressing the propagation of frequencies which promote fluorescence. 
	
	Resulting optimal predictions at each excitation wavelength are presented in table \ref{tab:resultsAll}; optimal predicted versus actual concentrations for glucose, excited at 488 nm, are depicted in figure \ref{fig:All_1to1} panel C.
	
	Solutions in this set were constructed with diabetes detection in mind and, as such, predictions must fall within 20\% of reference values above 3.89 mM (70 mg/dL) in order to be considered clinically accurate \cite{Clarke1987}. Thus, we define here the minimum permissible accuracy to be 0.778 mM. We exceed this accuracy by a wide margin when samples are measured using 488 nm, achieving 0.14 mM. When exciting the sample with 633 nm or 785 nm, however, accuracy falls short of this goal. 	
	
	\subsection{Waveguide Selection and Influence on Performance}
	
	Compatibility with our microfluidic system necessitates that the waveguide maintains light-guiding once immersed in fluid. \Glspl{ac:tct} immediately satisfy this requirement, suiting them for use if the facilities to prepare a waveguide for immersion are not available. Their sub-optimal scattering collection efficiency is easily compensated for with circulation and exposure times in excess of several hours. Over the same time scale in a static solution, laser heating evaporates the sample from the focal volume, compromises coupling, and introduces thermal effects such that the signal we observe is not representative of the bulk solution (figure \ref{fig:tempGlucose}). These thermal effects are mitigated with our closed-system continuous-flow optofluidics, as the fluid under investigation within the waveguide core is continually being refreshed and the thermal load is dispersed.
	
	\Glspl{ac:hcpcf} outperform \glspl{ac:tct} with regards to accuracy, however we demonstrate that difference in performance is largely mitigated using our algorithm for quantification, rather than peak fitting, the traditional standard. We present three different R$^2$ metrics to validate this claim: 1) correlation between concentration and peak height at the 1127 cm$^{-1}$ glucose Raman mode, 2) correlation between true versus predicted concentrations of glucose in all cases where statistical significance was identified by our algorithm, and 3) correlation between true and predicted concentrations of glucose in optimal statistically significant preprocessing methods. Results are presented in table \ref{tab:R2_methods}. Using the traditional peak fit, an R$^2$ difference of 0.11 is observed. This falls to a mere 0.02 once our algorithm is introduced.

	\subsection{Statistical Optimization of Principal Component Selection Models}
	
	The machine-learning methods which are utilized in this paper are an extension of those in which we propose that an F-test may be used to indicate whether a preprocessing method will yield a robust model, and to simultaneously select the optimal {\gls{ac:pc}} {\cite{Storey2019}}. Furthering those efforts, the effect of the indicating parameter which selects the optimal \glspl{ac:pc} for a predictive model cannot be negated. In this section we statistically optimize the indicating parameter for \gls{ac:pc} selection relative to the current indicating parameter, as it is used in previous work.	
	
	Ten indicative parameters which may act as markers to select the optimal \gls{ac:pc} were identified; a list of these parameters is included in table \ref{tab:indicatingVariables}. Forming a linear model between predictor (value associated with the indicating parameter) and response (\gls{ac:rss}), we use the model's R$^2$ value and median error to select candidates for optimization. Amongst all linear models of potential factors we observe R$^2$ = 0.23 $\pm$ 0.10 (median $\pm$ stdev) and median error 1.61 $\pm$ 0.37. The indicating parameter which simultaneously maximized R$^2$ and minimized median error for all data sets included in this paper was the product of five different factors, indicating that a non-linear solution may be optimal. Using the numbering convention listed in supplementary table \ref{tab:indicatingVariables}, the optimal parameter is 3$\times$4$\times$5$\times$7$\times$8. Performance of the optimal parameter relative to the original indicating parameter (used in \cite{Storey2019} and the data presented in figure \ref{fig:All_1to1} and table \ref{tab:resultsAll}) is included in table \ref{tab:R2Improve}, where it is apparent that the optimized parameter has greatly increased R$^2$ and reduced the median error. 
	
	\begin{table}[!t]
		\caption{R$^2$ correlations between predictor method and predicted values for concentrations of glucose in an artificial tear solution; our statistical preprocessing method is compared to estimations calculated by peak height at 1127 cm$^{-1}$, in either a \acrfull{ac:tct} or \acrfull{ac:hcpcf}.}
		\label{tab:R2_methods}
		\begin{center}
		\begin{tabular}{>{\centering\arraybackslash}m{4cm}  >{\centering\arraybackslash}m{1.5cm}  >{\centering\arraybackslash}m{1.5cm} }
			\textbf{Method}				& \textbf{TCT}	& \textbf{HC-PCF}\\
		  	\hline
		  	Peak Height				& 0.58	& 0.69\\
		  	All Significant			& 0.84	& 0.86\\
		  	Optimally Significant	& 0.95	& 0.99\\
		\end{tabular}
		\end{center}
	\end{table}

	\begin{table}[!t]
		\caption{Linear model correlations for the original Machine-Learning (ML) indicating parameter from \cite{Storey2019}, versus the optimized parameter from this paper.}
		\label{tab:R2Improve}
		\begin{center}
			\begin{tabular}{ >{\centering\arraybackslash}m{4cm}  >{\centering\arraybackslash}m{1.5cm}  >{\centering\arraybackslash}m{1.5cm} } 
				\hline
				\textbf{Indicating Parameter} & \textbf{R$^2$} & \textbf{Median Error} \\  
				\hline 
				Original, from \cite{Storey2019}		& 0.03				& 1.72\\
				ML Optimized		& 0.26				& 1.09\\
				\hline
			\end{tabular}
		\end{center}
	\end{table}
	
	This optimized indicating parameter is validated using the samples from this study which represent the closest approximation to real biofluids: glucose in an artificial whole blood. Results are summarized in table \ref{tab:PredError}. 

\section{Discussion}

	Non-invasive diagnostics via biofluid analysis in healthcare applications has undergone steady progress in recent years. Paper-based devices, smartphone-compatible microscopy systems, and wearable biosensors all offer rapid results outside of a laboratory setting, making healthcare more accessible by reducing reliance on specialized facilities and personnel \cite{Li2021}. These devices are highly specific, however, requiring reconfiguration for each detected analyte. Even label-free optical techniques, such as Surface-Plasmon Resonance or Mach-Zehnder Interferometers, require functionalized sensor surfaces which are specific to the target \cite{FernandezGavela2016}. Although some of these methods demonstrate sensitivity greater than that which we have shown here (a wearable smartphone-based biosensor has demonstrated a limit of detection of 35 $\mu$M glucose in sweat, for example), their utility is limited by the need to reconfigure \cite{Xiao2019}. This is the barrier which we address here with Raman spectroscopy and machine-learning. The use of Raman-based chemometrics for biofluid-based healthcare applications offers an alternative for label-free analysis with zero functionalization, and great strides have been made in recent years to increase reliability and robustness of predictive measurements with machine-learning. With regards to glucose detection in solution via Raman, we present a summary of historical related literature in table \ref{tab:otherSensitivities}, where we define a Figure of Merit (FOM) as the product of laser power and glucose sensitivity. We wish to minimize both of these parameters: low laser power decreases damage to biological samples and increases the likelihood that the technology can reasonably be implemented, and minimal sensitivity increases accuracy for confident diagnostics. We achieve here a reduction in the necessary laser power by two orders of magnitude for non-\gls{ac:sers} characterization, and a heightened sensitivity by two orders of magnitude for glucose quantification in solutions of high turbidity.
	
	The analysis system described here provides an enhanced Raman signal collection and vastly improved algorithms for portable, pre-treatment-free, reliable diagnostics in biofluids, without the need for \gls{ac:sers}. This has the potential to streamline diagnostics using Raman without the need to carry out any sample preparation, particularly for whole blood. Here we shall discuss the novelty and impact which each component in our system provides.
	
	The microfluidic device ensures that a given biofluid does not interact with potential environmental contaminants and does not experience thermal effects which may alter or obscure the Raman spectrum, thereby affecting diagnostic efficacy. The microfluidic configuration also eliminates evaporation, which ensures that optimal optical coupling is maintained during prolonged measurements. Environmental confinement also contributes to operator safety if the sample is may be pathogenic or toxic. To the best of our knowledge this is the first instance of a fully-contained microfluidic system with continuous circulation and an integrated waveguide for enhanced Raman spectroscopy.
	
	\Glspl{ac:lcw} allow us to achieve enhanced Raman collection efficiency which permits label-free non-invasive detection and monitoring. Traditional Raman is limited to collect a scattering signal from the laser pump's focal volume; the captured signal can be enhanced via larger \gls{ac:na} objectives to gather from a greater solid angle, but this comes at the expense of depth of focus and thereby the number of molecules in the focal volume. Confining both laser pump and sample in the core of a waveguide bypasses this \gls{ac:na} tradeoff, as molecules now interact with the pump along the entire waveguide length, increasing the number of scattering events and thereby the Raman signal \cite{Mak2013}. The enhanced collection efficiency which we achieve due to fully-contained \glspl{ac:lcw} allows us to integrate near-indefinitely, providing sensitivity beyond that needed to simply detect the presence of relevant Raman modes and bypassing the levels of sensitivity previously achieved using bare \glspl{ac:lcw} \cite{Qi2007a, Mak2013}. At the same time, circulation inhibits crystallization and analyte deposition during prolonged integration, ensuring that the composition which we observe is true to the sample's chemical makeup. With these novel modifications we can ensure accurate monitoring of minute changes between specimens.
	
	\begin{table}[!t]
		\caption{Machine-Learning prediction error for statistically significant preprocessing methods, acting on a solution of glucose in intralipid, in which the optimized indicating parameter is used to determine the principal component. Results may be directly compared to table \ref{tab:resultsAll} in which the original indicating parameter from \cite{Storey2019} is used.}
		\label{tab:PredError}
		\begin{center}
			\begin{tabular}{ >{\centering\arraybackslash}m{2cm}  >{\centering\arraybackslash}m{2.5cm}  >{\centering\arraybackslash}m{2.5cm} } 
				\hline
				\textbf{Excitation Wavelength} & \textbf{Significant (mM)} & \textbf{Optimal (mM)} \\  
				\hline 
				488 nm & 0.554 $\pm$ 0.483 & 0.0739 $\pm$ 0.307 \\ 
				633 nm & 42.4 $\pm$ 53.6 & 19.4 $\pm$ 29.4 \\ 
				785 nm & 15.9 $\pm$ 28.8 & 7.45 $\pm$ 24.2 \\ 
				\hline
			\end{tabular}
		\end{center}
	\end{table}

	\begin{table}[!t]
		\begin{threeparttable}
		\caption{Related literature: machine-learning predictions of glucose in solution. The product of laser power and sensitivity is presented as a figure of merit (FOM), as we wish to minimize these parameters, independent of integration time}
		\label{tab:otherSensitivities}
		\begin{center}
			\begin{tabular}{ >{\centering\arraybackslash}m{1.25cm}  >{\centering\arraybackslash}m{1.25cm}  >{\centering\arraybackslash}m{1.25cm} >{\centering\arraybackslash}m{1.25cm} >{\centering\arraybackslash}m{1.25cm}} 
				\hline
				\textbf{Reference} & \textbf{Laser Power (mW)} & \textbf{Integration Time (s)} & \textbf{Glucose Sensitivity (mM)} & \textbf{FOM (mW*mM)}\\  
				\hline 
				\cite{Berger1997}\tnote{*} 						& 150 	& 300 	& 3.6	& 540\\ 
				\cite{Berger1999}\tnote{\S} 					& 250 	& 60 	& 1.44	& 360\\
				\cite{Shafer-Peltier2003}\tnote{\S $\dagger$} 	& 1.25 	& 30 	& 1.8	& 2.25\\
				\cite{Lambert2005} \tnote{\S}					& 100 	& 3 	& 0.02	& 2\\
				\cite{Qi2007a}\tnote{\S}						& 160 	& 3 	& 0.49	& 78.4\\ 
				\cite{Storey2019}\tnote{\S $\ddagger$} 			& 17 	& 240 	& 0.22	& 3.74\\ 
				This Work\tnote{* $\ddagger$}					& 3.5 	& 20 	& 0.071	& 0.25\\
				\hline
			\end{tabular}
			\begin{tablenotes}\footnotesize
			\item[*] Sample of high turbidity
			\item[\S] Sample of low turbidity
			\item[$\dagger$] Surface-Enhanced Raman Spectroscopy
			\item[$\ddagger$] Our Work
			\end{tablenotes}
		\end{center}
		\end{threeparttable}
	\end{table}	
	
	These enhanced Raman collection methods achieve both sensitivity and accuracy for compounds which are difficult to characterize with traditional Raman spectroscopy, such as pre-treatment-free quantitfication of glucose in tears and whole blood \cite{Shafer-Peltier2003}. In a human tear phantom we achieve 0.274 mM versus the clinical accuracy limitation of 0.278 mM; in a human whole-blood phantom we achieve 0.14 mM versus clinical accuracy threshold of 0.778 mM \cite{VanHaeringen1977}. Accurate non-invasive glucose concentration measurements using photonic methods have, in the past, required significant sample and reagent preparation, such as \gls{ac:sers}. This is not feasible where specialized facilities are impractical to implement. In this study we have exceeded aqueous glucose errors-of-detection previously reported using \gls{ac:sers} (1.8 mM) by a significant margin, achieving 0.14 mM \cite{Shafer-Peltier2003}. Our spectrum collection methods demonstrate stability and reproducibility during lengthy acquisition times, further advocating the use of \gls{ac:lcw} Raman over \gls{ac:sers} in this application. 
	
	The reproducibility provided by our microfluidic design does not compensate for inherent signal variance between biological samples - this issue is addressed with our analytical techniques. Our user-independent machine-learning algorithm sufficiently identifies optimal preprocessing and regression methods to compensate for variance and yield reliable accurate diagnostics for a range of biomarkers, such as lysozyme in artificial human tear samples. The accuracies we achieve, assisted by the modified indicating parameter presented here, could readily aid diagnostics for \gls{ac:hsv}, Sj{\"o}gren's syndrome, and kerato-conjunctivitis \cite{Eylan1977}. Existing published studies which quantify lysozyme in aqueous fluids via \gls{ac:dcdrs} have achieved of errors-of-detection on the order of 28.6 $\mu$M or lower, similar in magnitude to our own \cite{Filik2007,Zhang2003}. \Gls{ac:dcdrs} has shown to be advantageous over \gls{ac:sers} in terms of reproducibility and stability, but not in magnitude of enhancement \cite{Zhang2003}. \Gls{ac:sers} can provide an enhancement upwards of 9 orders of magnitude versus unassisted Raman spectroscopy of aqueous samples, but the sample must be altered with nanostructures which renders it unusable for future study \cite{Mak2013}. Our methods achieve the reproducibility of \gls{ac:dcdrs} and spectrum enhancement of \gls{ac:sers} without the sample processing requirements.
	
	While this study has focused on two particular biofluids, performance can reasonably be extrapolated to other biological fluids which possess similar light-scattering properties, such as urine. Glucose levels in human urine between healthy and diabetic patients may vary from 2.78 mM to over 5.55 mM; by reasoning that tears and urine have similar turbidity, we can extrapolate our results from a human tear phantom to measure glucose with a sensitivity of 0.2 mM \cite{Bruen2017}. Glucose concentrations range from 0.01 mM to 5 mM in other biofluids such as sweat, saliva, and ocular fluids. Further refinement is necessary for our system to achieve confident glucose predictions on this scale; potential solutions include further extending exposure times. When considering alternate biofluids for constituent analysis, it is crucial to be aware of differences in light-matter interactions. Certain wavelengths may prompt a strong fluorescent response, but an appropriate pairing of excitation wavelength and photonic bandgap \gls{ac:lcw} (both of which our microfluidic device can readily adapt) promotes optimum characterization.
	
	To extend this device into field-deployable applications, two particular limitations must be addressed: with regards to applicability towards human fluids, and to the true limitation of analyte sensitivity. All artificial human biofluids in experiments described here are prepared from common stock solutions with a limited variety of constituents. In addition, we have normalized spectrum collection conditions across each data set. As such, the variances in our spectra do not reflect the range which we should expect from human biological samples. These experiments are a necessary foundation prior to testing with human biofluids; future implementations of this device will require testing on larger spectrum sets which do not come from the same data distribution, and which have a greater degree of variation in both measurement conditions and presence of other substances which may complicate the Raman spectrum of the analytes of interest. With regards to sensitivity, we argue that this does not reflect a limitation of our device. Artificial tear glucose sensitivity in our experiments is 1.4 mM; expected glucose concentrations in human tears vary substantially as a function of collection method and our experiments were designed to comply with one reported range. Enhanced sensitivities for alternate expected glucose ranges may be obtained by extending exposure times beyond those which we report. We have demonstrated system stability for acquisition times between three minutes and two hours on a tabletop vibration-isolated Raman system, with no changes in coupling into the \gls{ac:lcw} due to sample pumping around the fiber. Unwanted background and environmental interference is prevalent in portable Raman systems; our configuration is ideal for this scenario \cite{Capitan-Vallvey2011}. We expect a straightforward extension to interface the device described here with a portable spectrometer, satisfying the desire for portability and pre-treatment-free spectrum collection.
	
	Reliable and reproducible diagnostics is achieved via our algorithm, successfully automating selection of the optimal number of \glspl{ac:pc} for a model and eliminating unsuitable preprocessing treatments. This achieves a tremendous reduction in the number of decisions which require human intervention. Our artificial tears data set contains 40 spectra, 1 which acts as the unknown spectrum and 39 of which form a cross-validation training data set. This set forms a model containing up to 37 \glspl{ac:pc} for each possible preprocessing method. If we assess 28 different treatments, this results in 1036 possible unique predictive models. If even one preprocessing method is identified as unsuitable we reduce the number of decisions which require manual intervention from 1036 to just 27, eliminating 97.3\% of cases. The architecture of the algorithm developed here lends itself to complete automation; continuous improvements in accuracy and sensitivity will follow as larger datasets for machine-learning are made available, thereby bringing the manual intervention to zero for fully automated diagnostics.	

\section*{Supporting Information}
	Additional figures and tables are included in a separate supplementary information document. 

\section*{Acknowledgment}
	This work was supported by a Natural Sciences and Engineering Research Council of Canada (NSERC) Discovery Grant.

\ifCLASSOPTIONcaptionsoff
  \newpage
\fi



\bibliographystyle{IEEEtran}
\bibliography{MainBib}

\end{document}


\title{Enhanced Sensitivity for Quantifying Disease Markers via Raman and Machine-Learning of Circulating Biofluids in Optofluidic Chips\\\textbf{Supporting Information}}


\author{Emily~E.~Storey,
	    Duxuan~Wu,
	    Amr~S.~Helmy,~\IEEEmembership{Senior Member,~IEEE,}

\thanks{E.E. Storey and A.S. Helmy are with the Department of Electrical and Computer Engineering, University of Toronto, Toronto, Ontario, Canada.}
\thanks{Corresponding Author E-mail a.helmy@utoronto.ca}%
\thanks{D. Wu was with the Department of Electrical \& Computer Engineering, University of Toronto, Toronto, Ontario, Canada.}}

\IEEEoverridecommandlockouts
\IEEEpubid{\makebox[\columnwidth]{\begin{minipage}{\columnwidth}\copyright2021 IEEE. Personal use of this material is permitted. Permission from IEEE must be obtained for all other uses, in any current or future media, including reprinting/republishing this material for advertising or promotional purposes, creating new collective works, for resale or redistribution to servers or lists, or reuse of any copyrighted component of this work in other works.\hfill\end{minipage}} \hspace{\columnsep}\makebox[\columnwidth]{ }}
\maketitle
\IEEEpubidadjcol
\IEEEpubidadjcol


\begin{figure}[!t]
	\centering
	\includegraphics[width=\linewidth]{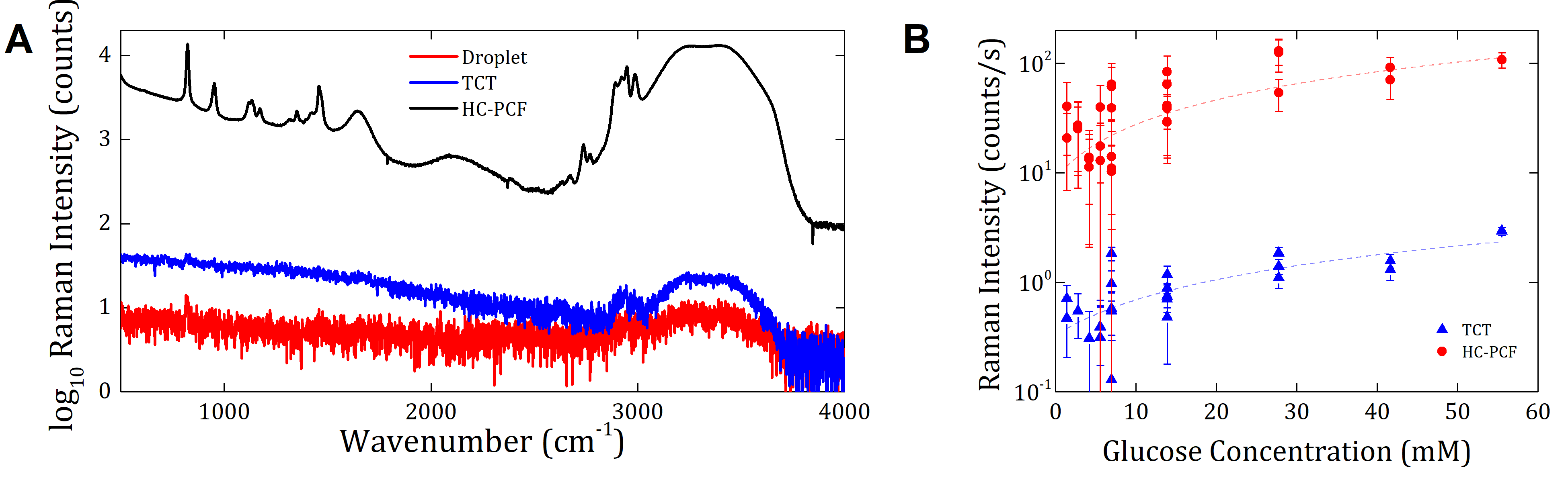}
	\caption{Raman spectrum enhancement method comparison. (A): Spectra are collected with a 1 second exposure and 5 acquisitions from a 10\% Isopropyl Alcohol and 90\% deionized water solution. Signal intensities are plotted between Raman spectra obtained from a droplet directly, versus the enhancement provided from a 4 cm length of Teflon Capillary Tube (TCT), and a 4 cm length of Hollow-Core Photonic Crystal Fiber (HC-PCF). (B): Raman intensity of the 1127 cm$^{-1}$ glucose peak normalized to acquisition exposure time as a function of glucose concentration. Mode intensity at 1127 cm$^{-1}$, as a function of glucose concentration measured using TCTs, is fitted to a linear equation with slope 0.04 and R$^2$ 0.58. The same treatment applied to HC-PCFs gives a slope of 1.87 and R$^2$ 0.69. Similar intensity spectra for this Raman mode can be collected approximately 50 times faster using a HC-PCF. }
	\label{fig:TCTvsDrop}
\end{figure}

\begin{figure}[!t]
	\centering
	\includegraphics[width=\linewidth]{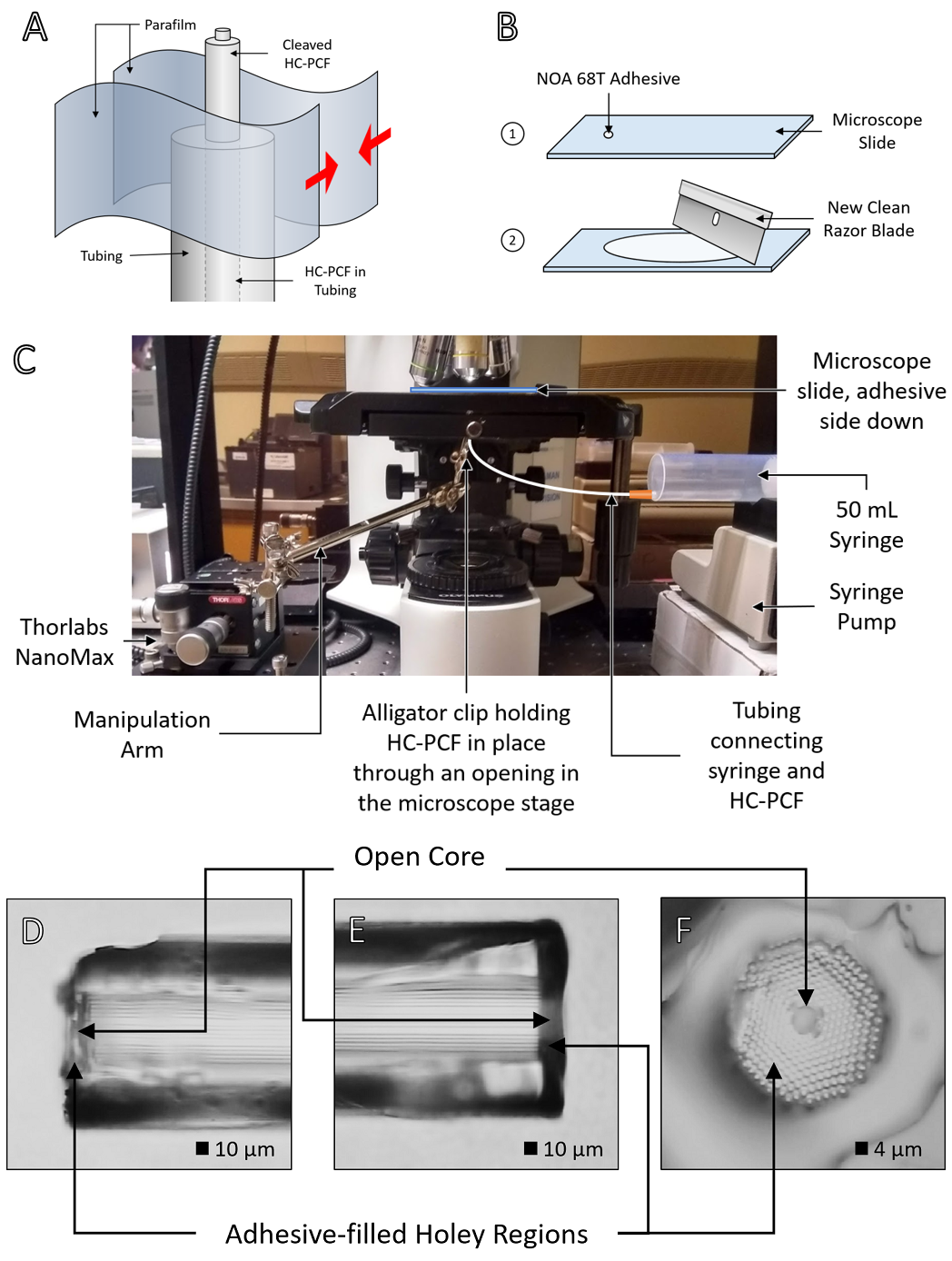}
	\caption{The PATTi two-ended Hollow-Core Photonic Crystal Fiber (HC-PCF) preparation method. (A): Cleaved fiber is partially inserted into tubing. Parafilm is applied around the fiber-tubing junction by pinching in the direction of the red arrows. (B): Adhesive (Norland Optical Adhesives NOA68T) is prepared by streaking a thin film on a microscope slide, then placed face-down on the microscope stage. (C): Fiber is clipped to a micromanipulation arm for alignment with microscope. Tubing attached to the fiber is connected to a syringe which delivers constant air pressure, ensuring the central core remains open. Adhesive is applied to the fiber tip by briefly making contact between the facet and adhesive-streaked slide. Adhesive is cured with a ultraviolet bulb and the process is repeated for the second fiber tip. (D): Side view of the first sealed HC-PCF end, with open core and adhesive filling a depth less than 10 $\mu$m in the holey region. (E): Side view of the second sealed end. Adhesive fills a length of 20 $\mu$m. The open core is not immediately apparent. (F): Facet view of the second sealed end. The open core is now evident.}
	\label{fig:supp-patti}
\end{figure}

\begin{table*}[!t]
	\centering
	\caption{Alternate indicating variables for optimal \gls{ac:pc} selection. To optimize performance, a linear model between predictor (value associated with the indicating parameter) and response (Residual Sum of Squares) is formed and the R$^2$ value and median error are used to select candidates for optimization.}
	\begin{tabular}{c m{8cm}}
		\hline
		\textbf{Number}	& \textbf{Description}\\
		\hline
		1				& Mean of mean differences in \gls{ac:press} value at each \gls{ac:pc}\\
		2				& Median of mean differences in \gls{ac:press} value at each \gls{ac:pc}\\
		3				& Standard deviation of mean differences in \gls{ac:press} value at each \gls{ac:pc}\\
		4				& Mean of p-values for each \gls{ac:pc}\\
		5				& Median of p-values for each \gls{ac:pc}\\
		6				& Standard deviation of p-values for each \gls{ac:pc}\\
		7				& Sum of \gls{ac:press} values for each \gls{ac:pc}\\
		8				& Mean of \gls{ac:press} values for each \gls{ac:pc}\\
		9				& Median of \gls{ac:press} values for each \gls{ac:pc}\\
		10				& Standard deviation of \gls{ac:press} values for each \gls{ac:pc}\\
		\hline
	\end{tabular}
	\label{tab:indicatingVariables}
\end{table*}	

\begin{table*}[!t]
	\centering
	\caption{Variation in median error of prediction for samples of glucose in an artificial whole blood solution. Erorrs are presented as median $\pm$ standard deviation.}
	\begin{tabular}{ >{\centering\arraybackslash}m{3cm}  >{\centering\arraybackslash}m{4cm}  >{\centering\arraybackslash}m{4cm}  >{\centering\arraybackslash}m{4cm} }
		\hline 
		\textbf{Excitation Wavelength} & \textbf{Optimal Significant Methods – Full Sample Range (mM)} & \textbf{Optimal Significant Methods – Full Sample Range (\% Error)} & \textbf{Optimal Significant Methods – 488 nm Sample Range (mM)} \\ 
		\hline
		488 nm & 0.0714 $\pm$ 0.214 & 6.43 $\pm$ 38.7 & 0.0714 $\pm$ 0.214 \\ 
		633 nm & 28.4 $\pm$ 27.5 & 700 $\pm$ 1956 & 1.28 $\pm$ 0.586 \\ 
		785 nm & 4.54 $\pm$ 23.3 & 136 $\pm$ 1241 & 0.744 $\pm$ 0.630 \\ 
		\hline 
	\end{tabular} 
	\label{tab:resultsByWavelength}
\end{table*}